\documentclass[aps,prl,twocolumn,longbibliography,noeprint,floatfix,superscriptaddress]{revtex4-2}

\usepackage{amsmath, amssymb}
\usepackage{mathtools}
\usepackage{lipsum}
\usepackage{times}
\usepackage{graphicx}
\usepackage{wasysym}
\usepackage{bm}
\usepackage{hyperref}
\usepackage{xspace}
\usepackage{siunitx}
\usepackage[usenames,dvipsnames]{xcolor}
\usepackage{soul}
\usepackage{centernot}
\usepackage{braket}
\usepackage{microtype}
\definecolor{myColor2}{rgb}{0.02,0.12,0.3}

\definecolor{myColor}{rgb}{0.02,0.12,0.7}
\definecolor{myciteColor}{rgb}{0.39,0.7,0.89}
\hypersetup{colorlinks=true, linkcolor=myColor, filecolor=myColor, urlcolor=myColor, citecolor=myColor, urlcolor=myColor}
\graphicspath{{./Figures/}}

\DeclareSIUnit{\nK}{\nano\kelvin}
\DeclareSIUnit{\um}{\micro\metre}
\DeclareSIUnit{\aB}{\emph{a}_0}
\DeclareSIUnit{\G}{G}

\newcommand{\kB}{k_{\text{B}}}

\newcommand{\g}{\tilde{g}}

\newcommand{\kD}{k_{\text{D}}}
\newcommand{\kF}{k_{\text{F}}}

\newcommand{\kx}{k_{\xi}}

\begin{document}

\title{
Observation of an Inverse Turbulent-Wave Cascade in a Driven Quantum Gas
}
\author{Andrey Karailiev}
\affiliation{Cavendish Laboratory, University of Cambridge, J. J. Thomson Avenue, Cambridge CB3 0HE, United Kingdom}
\author{Martin Gazo}
\affiliation{Cavendish Laboratory, University of Cambridge, J. J. Thomson Avenue, Cambridge CB3 0HE, United Kingdom}
\author{Maciej Ga\l ka}
\affiliation{Cavendish Laboratory, University of Cambridge, J. J. Thomson Avenue, Cambridge CB3 0HE, United Kingdom}
\affiliation{Physikalisches Institut der Universität Heidelberg,
Im Neuenheimer Feld 226, 69120 Heidelberg, Germany}
\author{Christoph Eigen}
\affiliation{Cavendish Laboratory, University of Cambridge, J. J. Thomson Avenue, Cambridge CB3 0HE, United Kingdom}
\author{Tanish Satoor}
\affiliation{Cavendish Laboratory, University of Cambridge, J. J. Thomson Avenue, Cambridge CB3 0HE, United Kingdom}
\author{Zoran Hadzibabic}
\affiliation{Cavendish Laboratory, University of Cambridge, J. J. Thomson Avenue, Cambridge CB3 0HE, United Kingdom}

\begin{abstract}

We observe an inverse turbulent-wave cascade, from small to large lengthscales, in a driven homogeneous 2D Bose gas. Starting with an equilibrium condensate, we drive the gas isotropically on a lengthscale much smaller than its size, and observe a nonthermal population of modes with wavelengths larger than the drive one. At long drive times, the gas exhibits a steady nonthermal momentum distribution. At lengthscales increasing from the drive one to the system size, this distribution features in turn: (i) a power-law spectrum with an exponent close to the analytical result for a particle cascade in weak-wave turbulence, and (ii) a spectrum reminiscent of a nonthermal fixed point associated with universal coarsening in an isolated 2D gas. In further experiments, based on anisotropic driving, we reveal the complete qualitative picture of how the steady-state cascade forms.

\end{abstract}
\maketitle


\begin{figure}[t]
    \centering
\includegraphics[width=\columnwidth]{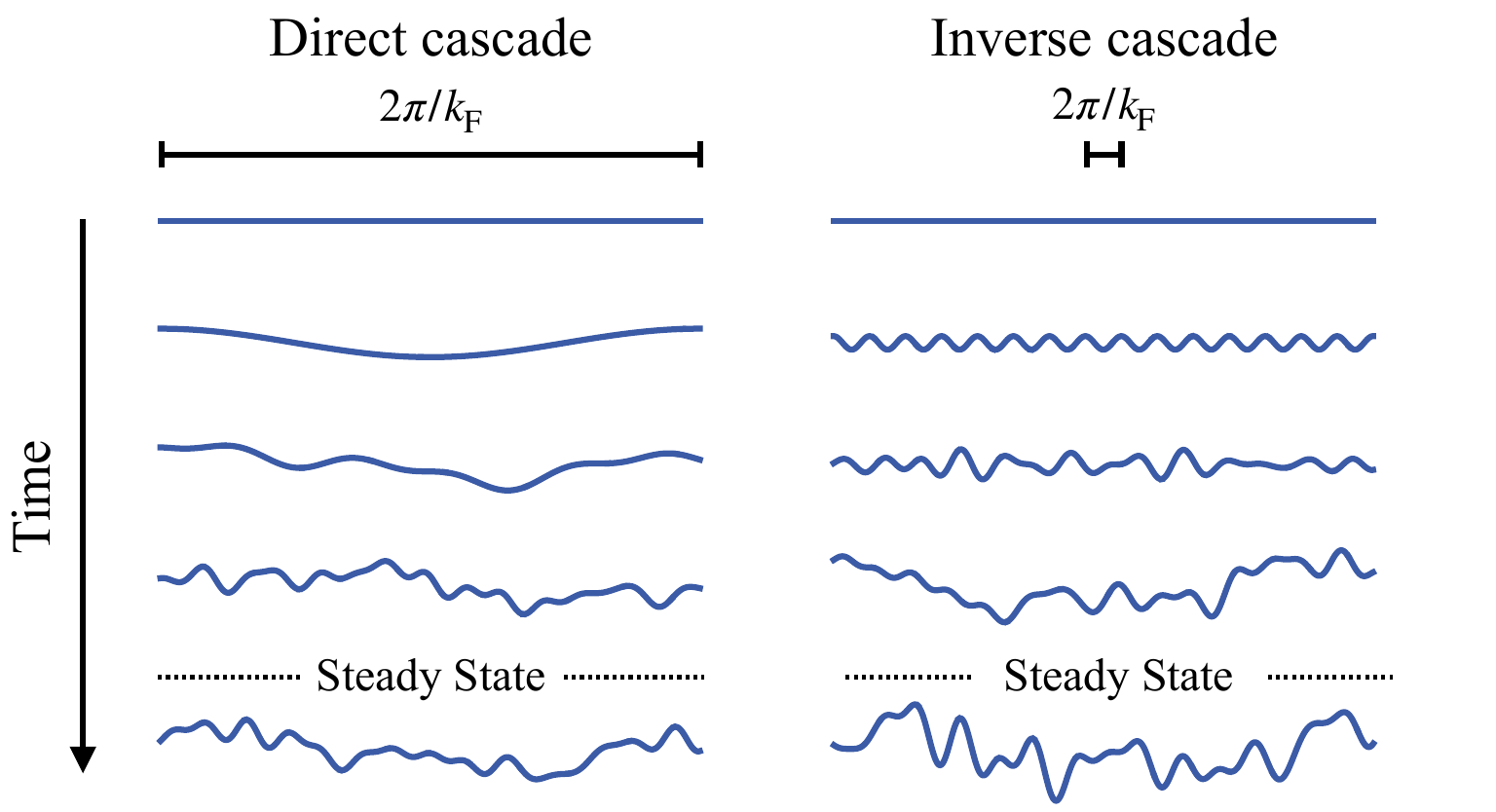}
\caption{
Real-space cartoons of direct and inverse cascades. The direct cascade corresponds to the transport of excitations from large to small lengthscales (small to large wavenumbers), and the inverse one to transport from small to large lengthscales. Here, starting at time $t=0$, the system is continuously driven at the wavenumber $\kF$; on the left $1/\kF$ is large and oscillations on ever-smaller lengthscales appear as the system evolves, while on the right $1/\kF$ is small and larger-scale oscillations appear. For weak-wave turbulence in a Bose gas, the direct cascade with an idealized sink (dissipation) at $\kD \rightarrow \infty$ corresponds to a scale-invariant energy flux in momentum space, without any particle flux. Similarly, the inverse cascade with a sink at $\kD \rightarrow 0$ corresponds to a pure particle flux, without any energy flux. In both cases, the quantitative signature of a steady-state cascade (established in finite time for a finite $\kD$) is a nonthermal power-law distribution of mode occupations.
}
 \label{fig:1}
\end{figure}


Richardson's direct cascade~\cite{Richardson:1922,Kolmogorov:1941,Obukhov:1941}, which transports excitations from large to small lengthscales, provides a canonical picture of turbulence and underpins its modern understanding in systems ranging from interplanetary plasmas to financial markets~\cite{Sorriso:2007,Ghashghaie:1996}. However, depending on the transport-conserved quantities, turbulence can also feature inverse cascades~\cite{Kraichnan:1967}, from small to large lengthscales (see Fig.~\ref{fig:1}). Such inverse cascades, like the direct ones, are naturally revealed in momentum space, and their key predicted signatures are nonthermal, power-law spectra of quantities such as energy or wave amplitude. For vortex-dominated turbulence, inverse cascades are seen~\cite{Paret:1997,Rutgers:1998,Young:2017,Gauthier:2019,Johnstone:2019,Panico:2023}, for example, in Jupiter's atmosphere~\cite{Young:2017} and Onsager's vortex clustering~\cite{Gauthier:2019,Johnstone:2019,Panico:2023}. For wave turbulence, they have been predicted for systems including gravitational waves in the early Universe~\cite{Galtier:2017,Galtier:2019} and magnetized plasmas in neutron stars~\cite{Biskamp:1996,Brandenburg:2020}. Experimentally, evidence for a decaying inverse wave cascade was observed in nonlinear optics~\cite{Bortolozzo:2009, Laurie:2012, Pierangeli:2016}, while evidence of a steady-state one was observed for surface waves on fluids, using spatially local measurements in the time domain~\cite{Deike:2011,Falcon:2020,Falcon:2022}, but it has remained an outstanding goal to observe a steady-state inverse wave cascade directly in momentum space.

Ultracold Bose gases present an ideal system for studies of wave turbulence on all relevant lengthscales and with well controlled driving and dissipation, as was previously demonstrated for a direct cascade in both three~\cite{Navon:2016,Navon:2019,Dogra:2023} and two dimensions~\cite{Galka:2022}. In this Letter we realize and study an inverse turbulent-wave cascade in a homogeneous 2D Bose gas~\cite{Chomaz:2015, Navon:2021}, by continuously driving it on a lengthscale much smaller than its size. Starting with an equilibrium condensate and isotropically exciting particles to a large wavenumber $\kF$, at long times we observe a stationary nonthermal momentum distribution $n_k$. For wavenumbers $k_{\xi} \lesssim k\lesssim\kF$, where $1/k_{\xi}$ is the healing length of the initial condensate, we observe a power-law spectrum $n_k \propto k^{-\gamma}$, with the exponent $\gamma=1.55(15)$ close to the analytical result $\gamma=4/3$ for a particle cascade in weak-wave turbulence (WWT) in a 2D Bose gas~\cite{Dyachenko:1992}. At lower $k$ this behavior breaks down, but $n_k$ also differs from that of the equilibrium condensate. Instead, it shows similarity to the spectrum associated with a nonthermal fixed point (NTFP)~\cite{Berges:2008} that characterizes coarsening in an isolated 2D Bose gas~\cite{Chantesana:2019, Gazo:2023}; this is consistent with the system effectively behaving as strongly interacting at large lengthscales. We also show that the salient features of the cascade survive for anisotropic (uniaxial) driving, and study the dynamics of the steady-state formation, observing qualitative agreement with theory~\cite{Semikoz:1995,Galtier:2000,Semisalov:2021,Zhu:2023b}. 

We start with a 2D quasi-pure $^{39}$K condensate of $5\times 10^4$ atoms in the lowest hyperfine state, confined in the $x$-$y$ plane by a circular optical box trap of radius $R=\textnormal{\SI{22}{\um}}$ and depth $U_{\rm D}=\kB\times \SI{150}{\nano\kelvin}$; the gas is confined to the plane by a harmonic trap with frequency $\omega_z/(2\pi)=\SI{1.5}{\kilo\hertz}$~\cite{Christodoulou:2021}. We set the $s$-wave scattering length $a$ to $30\,a_0$ (where $a_0$ is the Bohr radius) using the $402.7$-G Feshbach resonance~\cite{Etrych:2023}, so the dimensionless interaction parameter is $\g=a\sqrt{8\pi m\omega_z/\hbar}=0.02$ (where $m$ is the atom mass)~\cite{Hadzibabic:2011}.
The particle density is $n=\textnormal{\SI{33}{\um^{-2}}}$, the inverse healing length is $k_\xi=\sqrt{\tilde{g}n}\approx\textnormal{\SI{0.8}{\per\um}}$, and the trap depth sets the high-$k$ dissipation scale, $k_{\rm{D}}=\sqrt{2 m U_{\rm D}}/\hbar\approx\textnormal{\SI{5}{\per\um}}$~\cite{Navon:2019}. We use matter-wave focusing~\cite{Tung:2010,Gazo:2023} to measure the 2D momentum distributions $n_k$, which we normalize so that $\int n_k\, \mathrm{d}^2{\bf k} = 1$.

\begin{figure}[t]
\centering
\includegraphics[width=\columnwidth]{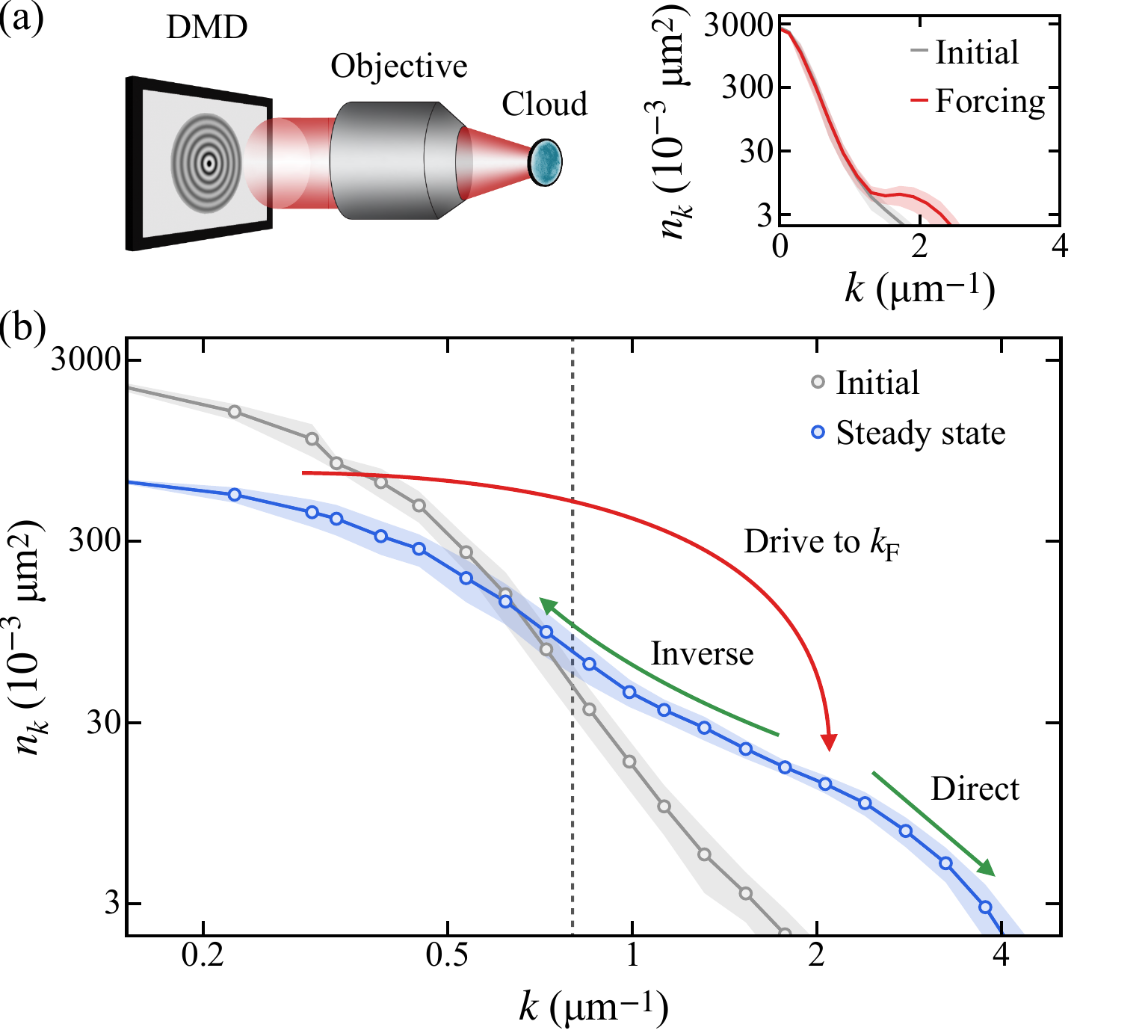}
\caption{
Forcing and steady state. (a) Left: we drive our box-trapped atomic cloud of radius $R=\textnormal{\SI{22}{\um}}$ isotropically at $\kF\approx\textnormal{\SI{2.1}{\per\um}} \gg 1/R$, using a digital micromirror device (DMD) to create a temporally oscillating potential~\cite{Supplementary}. Right: momentum distribution after two drive periods (red) shows excitations above the initial equilibrium state (gray). (b) Far-from-equilibrium steady state (blue), shown for $\omega t/(2\pi)=200$ and contrasted with the initial equilibrium distribution (gray); the dashed line indicates $k_\xi=\SI{0.8}{\per\um}$. In both panels, the lines and points show azimuthally-averaged data, and the shading indicates azimuthal variations (standard deviation). From hereon (Figs. 3 and 4) we show just the azimuthal averages.}
 \label{fig:2}
\end{figure}

As in theoretical studies~\cite{Dyachenko:1992,Nazarenko:2006,Nazarenko:2011,Zhu:2023a}, we drive the system isotropically at $|\textbf{k}|=\kF\gg1/R$; however, note that theorists typically consider particles injected into an initially empty system, whereas the particles we inject at $\kF$ initially come from the condensate and are at later times recirculated from the low-$k$ states. As illustrated in Fig.~$\ref{fig:2}$(a), we use a digital micromirror device (DMD) to project onto the atomic cloud a grayscaled~\cite{Gauthier:2016,YQZou:2021} radially symmetric driving potential, $U(r, t) = f(r)\sin(\omega t)$, where $f(r)$ is chosen to mimic the spatial structure of the excited mode~\cite{Supplementary}. The approximate radial period of $f(r)$ is \SI{3.0}{\um}, which sets $2\pi/\kF$, so $\kF=\textnormal{\SI{2.1}{\per\um}}$. Our chosen drive frequency, $\omega/(2\pi)=\SI{710}{\hertz}$, matches the Bogoliubov-dispersion $\omega({\kF})$~\cite{StrongDrivefn}.

At long drive times, $\omega t /(2\pi) \geq 75$, we observe a (quasi-) steady far-from-equilibrium $n_k$, such as shown in Fig.~\ref{fig:2}(b) for $\omega t/(2\pi)=200$. As indicated by the arrows, our driving at $\kF$ (red) results in both inverse and direct cascades (green).
Since our $\kD$ is not infinite, the direct cascade carries a nonzero particle flux~\cite{Navon:2019} and results in a slow reduction of the total atom number (by $30\%$ between $\omega t/(2\pi) = 75$ and $200$), but we find that the normalized $n_k$ remains essentially stationary. Here we focus on the inverse cascade, while the direct one was studied with low-$\kF$ driving in~\cite{Galka:2022}; in that case $n_k \propto k^{-2.9}$ was observed [indicated in Fig.~\ref{fig:2}(b) by the slope of the arrow].

\begin{figure}[t]
\centering
\includegraphics[width=\columnwidth]{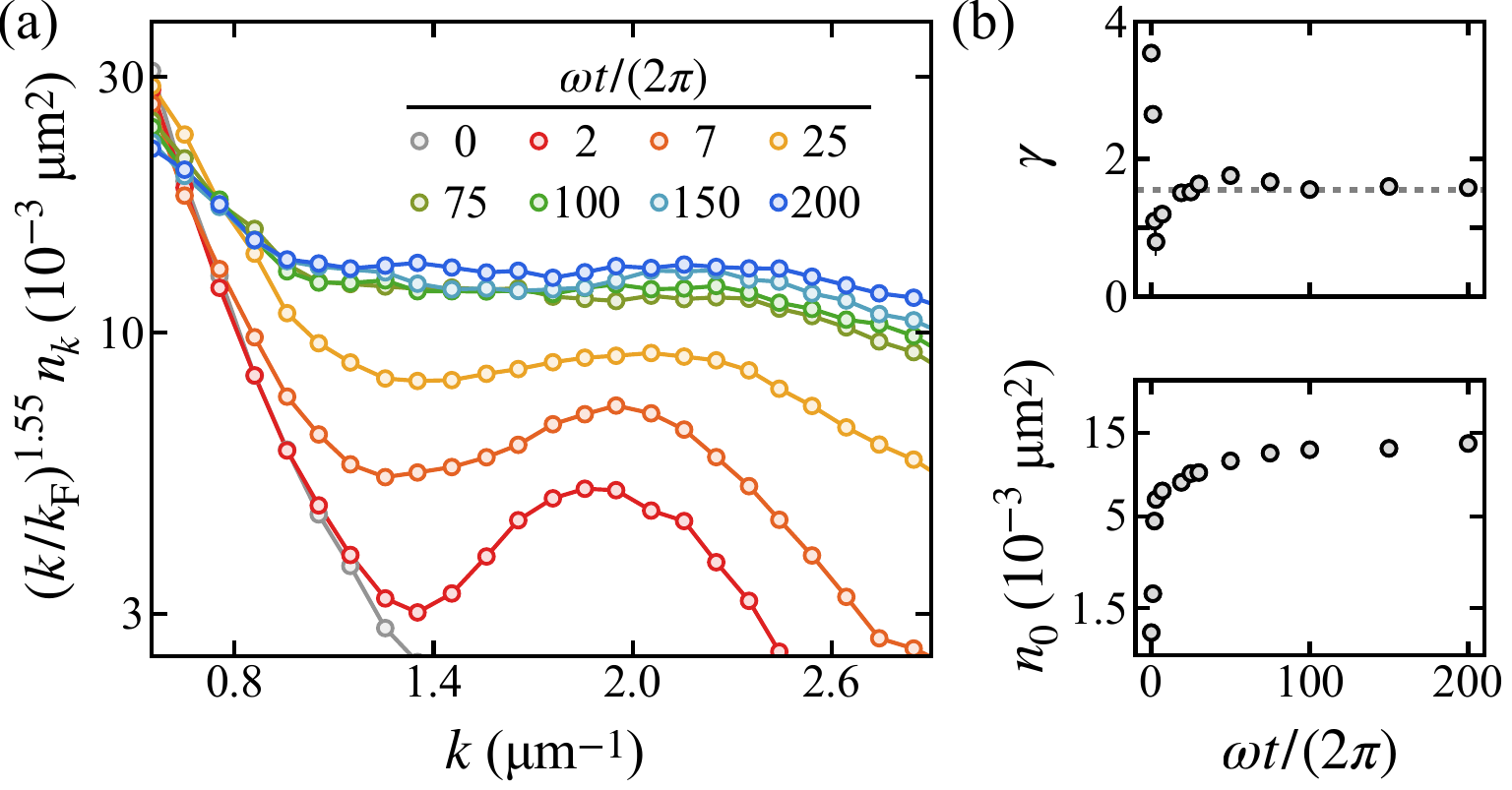}
    \caption{Steady-state $n_k$ at $k_\xi \lesssim k \lesssim \kF$. (a)
    The flatness of the compensated spectra, $(k/\kF)^{1.55}\,n_k$, shows that the nearly stationary long-time distributions are close to the power-law form $n_k = n_0 \, (k/\kF)^{-\gamma}$, with exponent $\gamma=1.55$. The analytical result for a particle cascade in WWT is $\gamma=4/3$. (b) Cascade amplitude, $n_0$, and exponent, $\gamma$, extracted from power-law fits of $n_k$ for $k=\textnormal{\SIrange{1.0}{2.1}{\per\um}}$; note that for $\omega t/(2\pi) < 50$ the distribution is not really a power law in this $k$-range. The dashed line shows $\gamma=1.55$. 
    }
 \label{fig:3}
\end{figure}

\begin{figure}[b]
\centering
\includegraphics[width=\columnwidth]{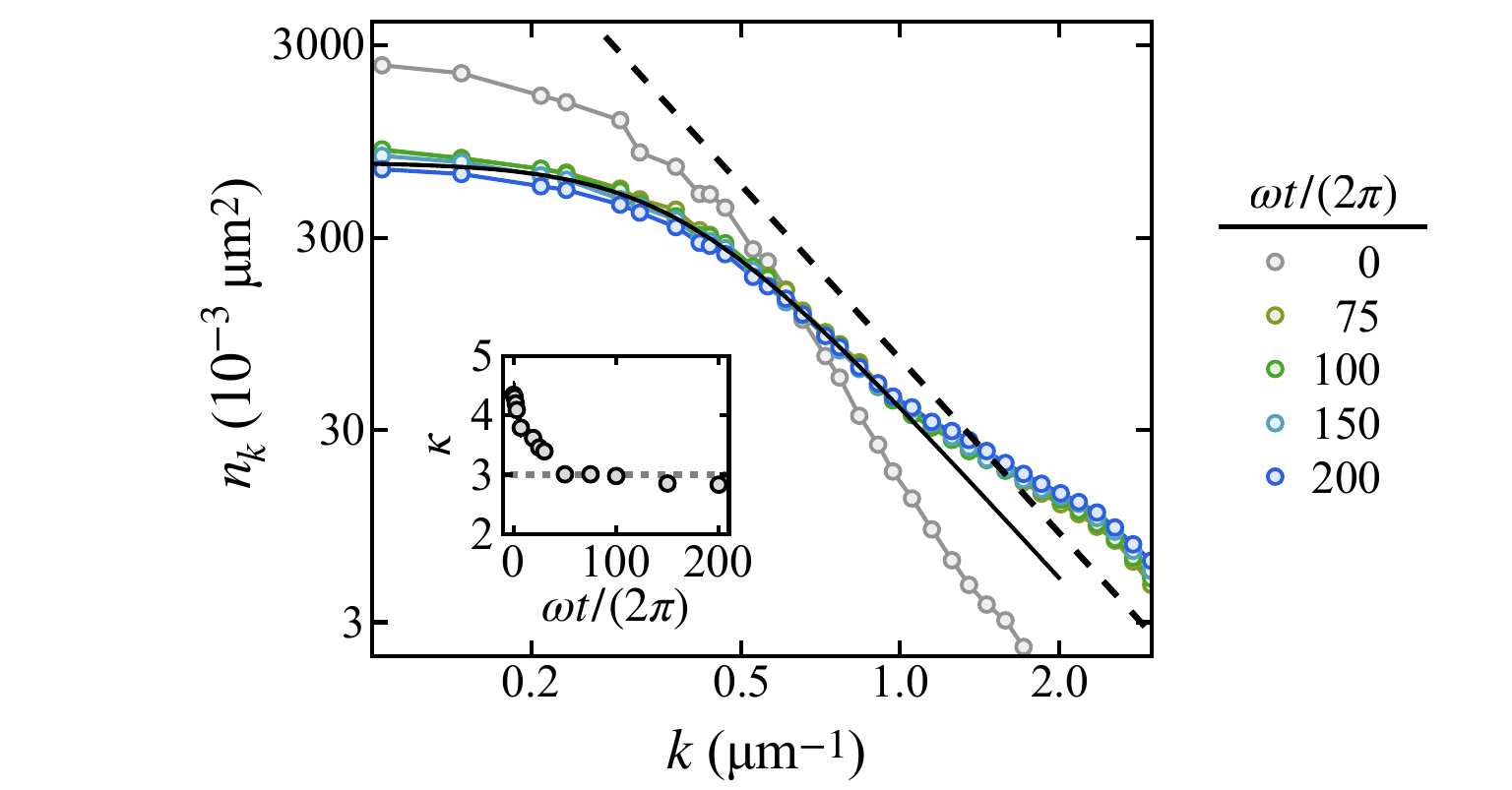}
\caption{
Steady-state $n_k$ at $k\lesssim k_{\xi}$. At long times, $n_k$ for $k\leq \SI{1}{\per\micro\metre}$ is captured well by $n_k\propto 1/(1+(k/k_0)^\kappa)$ with exponent $\kappa=3$ (solid black line; the dashed line shows the power-law $n_k \propto k^{-3}$). The same form of $n_k$, with the same $\kappa$, characterizes coarsening in an isolated 2D Bose gas, but in that case $k_0$ decreases algebraically in time, whereas here, under continuous driving, it remains constant ($\approx\textnormal{\SI{0.4}{\um^{-1}}}$). In this spectral range our $n_k$ is at all times captured reasonably well by the same functional form, with $\kappa$, fitted as a free parameter, evolving as shown in the inset.
}
 \label{fig:4}
\end{figure}

\begin{figure*}[t]
    \centering
\includegraphics[width=\textwidth]{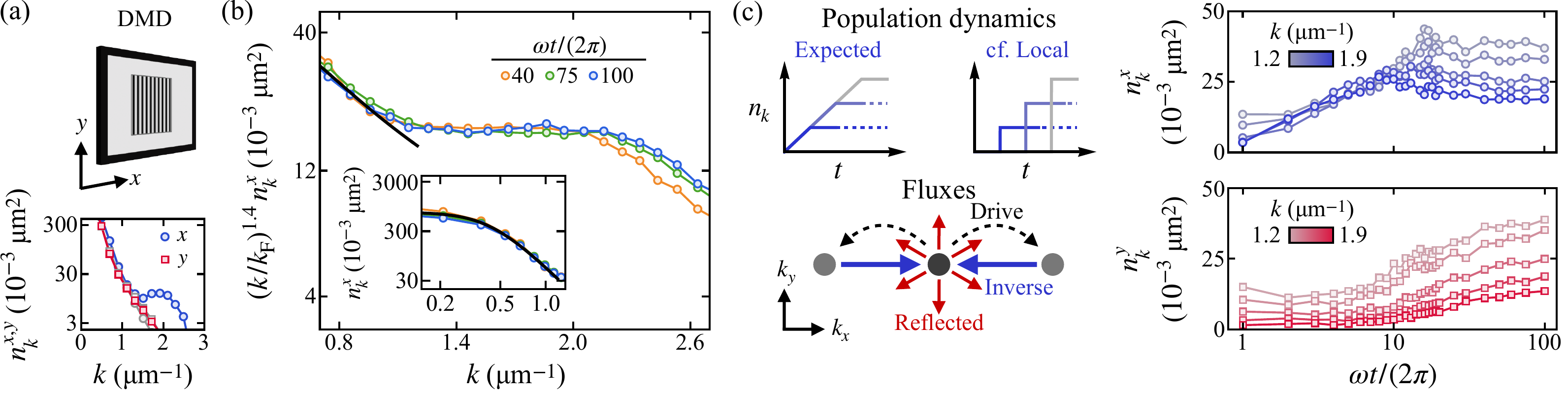}
\caption{
Steady state and cascade-formation dynamics for anisotropic driving. (a) Here we use a square box trap and change the DMD pattern to drive the gas uniaxially, at $(k_x, k_y)=(\pm \kF, 0)$, with the same $\kF$ as in Figs.~\ref{fig:2}-\ref{fig:4}. The bottom panel shows, for $\omega t/(2\pi) = 2$ (in color; cf. $ t = 0$ in gray), that excitations are created along $x$ and not along $y$. Here $n_k^{x}(k)$ (resp. $n_k^{y}(k)$) is the average population of modes with wavevectors within $\pi/8$ radians of the $x$ (resp. $y$) axis. (b) Along $x$ we observe essentially the same steady state as for isotropic driving, with $\gamma = 1.4(1)$ for $k_\xi \lesssim k \lesssim \kF$ and $\kappa =3$ (black line) at lower $k$; the inset shows a zoom-in on low $k$.  (c) Population dynamics along $x$ and $y$. The top cartoons show the expected non-local dynamics along the drive and, for comparison, what the dynamics would be if they were local; darker blue corresponds to higher $k$. The bottom cartoon shows the uniaxial drive to $(\pm \kF, 0)$, the inverse particle flux, and the reflected flux; for isotropic driving both fluxes are isotropic, while here we expect the inverse one to be uniaxial and the reflected one to be omnidirectional. Our $n_k^x$ data (top panel) are consistent with the expected non-local dynamics, while the $n_k^y$ data (bottom panel) reveal the delayed reflected flux.
}
 \label{fig:5}
\end{figure*}

In Fig.~\ref{fig:3} we focus on the spectral range $k_\xi \lesssim k \lesssim \kF$, where the equilibrium dispersion relation is particle-like. In Fig.~\ref{fig:3}(a) we show the evolution of $n_k$, which at long times is nearly stationary and shows the hallmark feature of a turbulent cascade: a power-law $n_k = n_0 \, (k/\kF)^{-\gamma}$. Our exponent $\gamma = 1.55(15)$ is close to $4/3$, the analytical result for a particle cascade in WWT theory for weakly interacting 2D Bose gases~\cite{Dyachenko:1992} (see also~\cite{Warmfn}); the discrepancy between the two could arise due to finite-size effects~\cite{Chantesana:2019,Galka:2022,Dogra:2023,Zhu:2023a,Martirosyan:2024}. In Fig.~\ref{fig:3}(b) we summarize the system evolution towards the steady state by plotting $n_0(t)$ and $\gamma(t)$ obtained from power-law fits of $n_k$.

At lower $k$, this behavior breaks down. Here, the initial condensate is replaced by a nonthermal distribution that is at long times captured well by $n_k \propto 1/(1+(k/k_0)^\kappa)$ with exponent $\kappa=3$ (see Fig.~$\ref{fig:4}$). Interestingly, the same form of $n_k$, with the same $\kappa$, characterizes the particle flow to low $k$ during coarsening in an isolated 2D Bose gas~\cite{Chantesana:2019, Gazo:2023}, which is understood in terms of the system being attracted to an NTFP~\cite{Berges:2008}. However, during coarsening $k_0$ decreases algebraically in time, whereas here, under continuous driving, the excitations are recirculated in $k$-space and $k_0$ is constant~\cite{ParticleRecircfn}. Tentatively, this can be interpreted as the system being stabilized near an NTFP. 

The exponents $\gamma$ and $\kappa$ are in fact predicted within the same theoretical framework of the momentum-space wave-kinetic equation, but in the weakly- and strongly-interacting limits, respectively~\cite{Chantesana:2019,Classicalfn}. Our observations are thus also consistent with the theoretical expectation that the system behaves as weakly interacting on lengthscales smaller than $1/\kx$, but as strongly interacting on larger lengthscales.

We now turn to experiments where we change the driving potential to further study the cascade-formation dynamics. Here we drive the gas along only one direction, $x$ [see Fig.~\ref{fig:5}(a)], and also switch to a square box trap (of side length \SI{30}{\micro\metre}), while keeping $n$, $k_\xi$, and $\kF$ the same as in Figs.~\ref{fig:2}-\ref{fig:4}. 

First, in Fig.~\ref{fig:5}(a) we show, for $\omega t/(2\pi) = 2$, that the injection of excitations into modes with wavevectors close to the $x$ axis is similar to before [see Fig.~\ref{fig:2}(a)], but the modes with wavevectors close to the $y$ axis are not directly excited by the drive. Here $n_k^x (k)$ denotes populations averaged over ${\bf k}$ vectors within $\pi/8$ radians of the $x$ axis (parallel to the drive), and  $n_k^y (k)$ denotes populations of modes with wavevectors within $\pi/8$ radians of the $y$ axis (perpendicular to the drive).

Next, in Fig.~\ref{fig:5}(b) we show that along the drive direction the long-time behavior is essentially the same as for the isotropic drive, which implies that the cascade dynamics are azimuthally local: for $k_\xi \lesssim k \lesssim \kF$ we observe a stationary power-law $n_k^x$, with $\gamma = 1.4(1)$, while $n_k^x$ at lower $k$ is captured well by the same form as in Fig.~\ref{fig:4}, with $\kappa =3$. Note that here the steady state is reached for $\omega t /(2 \pi) \approx 40$.

Finally, in Fig.~\ref{fig:5}(c) we study the cascade formation by looking at the dynamics of both $n_k^x$ and $n_k^y$; the cartoons outline the two key theoretical concepts that our data support. First, for isotropic driving, $n_k$ at different $k$ are expected to initially grow in unison~\cite{Semikoz:1995,Semisalov:2021}, which means that the inverse cascade, in contrast to the direct one~\cite{Galka:2022}, is not local in $k$-space (see top cartoons). Here, this expectation should apply (only) to the dynamics of $n_k^x$, parallel to the drive, and indeed matches our observations. Second, in theory, the inverse particle flux also results in a delayed counter-propagating `reflected' flux~\cite{Galtier:2000,Zhu:2023b}, which is formally required for the steady state to be established~\cite{Reflectfn}. We do not observe this flux directly for isotropic driving, but here it is revealed perpendicular to the drive, where there is no inverse flux (see bottom cartoon): the $n_k^y$ populations do grow, but only with a delay, after the growth of $n_k^x$  (for $k\gtrsim k_\xi$) is nearly complete. At long times, $n_k^y$ is also close to a power-law, approximately $\propto k^{-2.4}$ (not shown, see~\cite{Supplementary}), but we are not aware of any prediction for this purely reflected cascade.

In conclusion, by measuring directly in $k$-space and on all relevant lengthscales and timescales, we observe both the statistical properties of a steady-state inverse turbulent-wave cascade in a driven 2D Bose gas, and the key qualitative features of its formation dynamics. In the future, it would be interesting to measure the underlying cascade fluxes~\cite{Navon:2019} and test the expected self-similar evolution at pre-steady-state times~\cite{Semisalov:2021,Zhu:2023b}. Such studies would greatly benefit from an extension of the cascade spectral range (here limited by our optical resolution~\cite{Supplementary}), which could be increased by an order of magnitude using an optical lattice to drive the gas, capitalizing on our observation that uniaxial driving preserves the key features of the cascade. At low $k$, the stationary nonthermal spectrum we observe has a characteristic lengthscale ($1/k_0$), whose origin could be linked to the recently proposed concept of emergent lengthscales in nonlinear driven systems~\cite{deWit:2024}.

The supporting data for this Letter are available in the Apollo repository~\cite{Karailiev:2024-dataApollo}.

We thank Panagiotis Christodoulou and Nishant Dogra for early contributions and Gevorg Martirosyan, Jiří Etrych, Nir Navon, Jürgen Berges, and Aleksas Mazeliauskas for discussions and comments on the manuscript. This work was supported by EPSRC [Grant No.~EP/P009565/1], ERC (UniFlat), and STFC [Grant No.~ST/T006056/1]. M. Ga{\l}ka acknowledges support from Germany’s Excellence Strategy EXC2181/1-390900948 (Heidelberg Excellence Cluster STRUCTURES). Z.H. acknowledges support from the Royal Society Wolfson Fellowship.

%


\setcounter{figure}{0} 
\setcounter{equation}{0}
\renewcommand\theequation{S\arabic{equation}} 
\renewcommand\thefigure{S\arabic{figure}} 
\bigskip

\section*{SUPPLEMENTAL MATERIAL}

\subsection{DMD driving}

We use grayscaling on a digital micromirror device (DMD) to create our time-varying drive potentials. Our imaging resolution is approximately~\SI{1.5}{\micro\metre}, whereas a DMD pixel corresponds to \SI{0.22}{\um} in the atomic plane, which gives \mbox{$(1.5/0.22)^2 \approx 46$} grayscale levels~\cite{Gauthier:2016, YQZou:2021}. As in~\cite{YQZou:2021}, we use an error diffusion algorithm to convert continuous functions into binary masks. The largest $\kF$ we can achieve, of about \textnormal{\SI{2}{\per\um}}, is also set by our imaging resolution.

In the circular trap (with radius $R=\textnormal{\SI{22}{\um}}$) used for Figs.~\ref{fig:2}-\ref{fig:4}, radially symmetric excitation modes have the form $J_0(q_n r/R)$, where $J_0$ is the zeroth-order Bessel function of the first kind and $q_n$ its $n$\textsuperscript{th} extremum. In the square trap (with side length $L=\textnormal{\SI{30}{\um}}$) used for Fig.~\ref{fig:5}, we excite a mode of the form $\cos(n\pi (x+L/2)/L)$.

For the radial and axial driving, respectively, we choose $n=14$ and $21$, which in both cases corresponds to $\kF\approx\textnormal{\SI{2.1}{\per\um}}$. Note, however, that due to strong driving and imperfect mode-matching some neighboring modes could also be directly excited by the drive; in both cases the closest modes allowed by the symmetry of the drive have $k$ values that differ from our nominal $\kF$ by about $\pm\textnormal{\SI{0.2}{\per\um}}$.

\begin{figure}[b!]
    \centering
\includegraphics[width=\columnwidth]{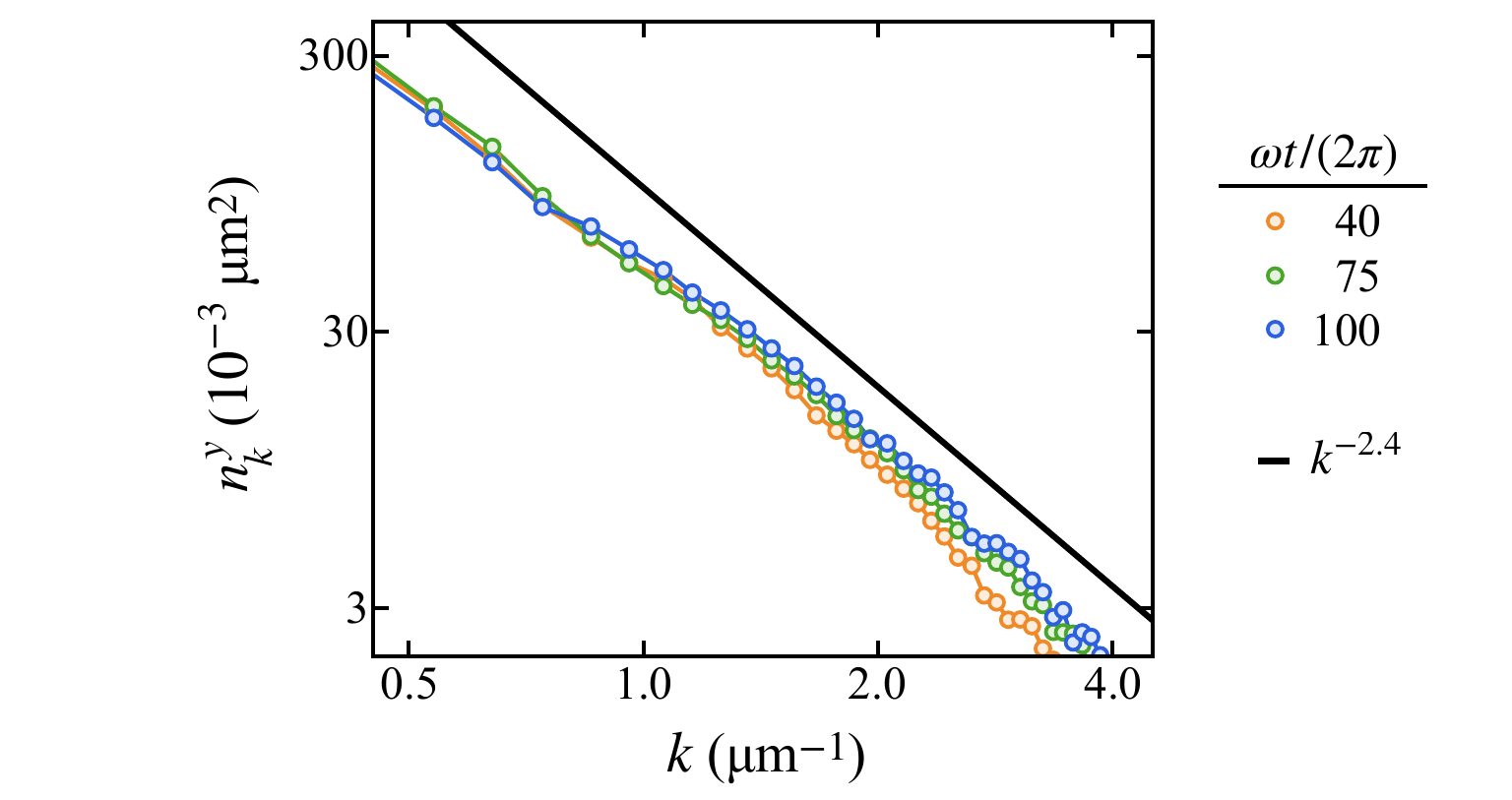}
\caption{Long-time momentum distributions $n_k^y$, perpendicular to the uniaxial drive. The black line corresponds to $n_k^y\propto k^{-2.4}$.}
 \label{fig:SI-axialky}
\end{figure}

\subsection{Additional panel for Fig.~5}

The uniaxial driving protocol used for Fig.~\ref{fig:5} results in anisotropic momentum distributions at long drive times. In Fig.~\ref{fig:SI-axialky} we show $n_k^y$ for the same times as in Fig.~\ref{fig:5}(b). The black line corresponds to $n_k^y\propto k^{-2.4}$.


\begin{thebibliography}{54}
\bibitem [{Str()}]{StrongDrivefn}%
  \BibitemOpen
  \href@noop {} {}\bibinfo {note} {For our strong drive, we observe efficient excitation up to about $\SI{120}{\hertz}$ away from $\SI{710}{\hertz}$.}\BibitemShut {Stop}%
\bibitem [{War()}]{Warmfn}%
  \BibitemOpen
  \href@noop {} {}\bibinfo {note} {We note that the validity of the result $\gamma=4/3$ was questioned by the authors themselves~\cite{Dyachenko:1992}, because it is formally not self-consistent. In 1D optical-wave turbulence the same type of result is also formally not self-consistent, but does agree quite well with the experiments~\cite{Laurie:2012}.}\BibitemShut {Stop}%
\bibitem [{Par()}]{ParticleRecircfn}%
  \BibitemOpen
  \href@noop {} {}\bibinfo {note} {The particle recirculation would be perfect for \mbox{$\kD/\kF \rightarrow \infty$}, such that the direct cascade would not carry away any particles. In our case, the normalized $n_k$ is stationary.}\BibitemShut {Stop}%
\bibitem [{Cla()}]{Classicalfn}%
  \BibitemOpen
  \href@noop {} {}\bibinfo {note} {This is a classical-field Gross--Pitaevskii model, which should be applicable when the mode occupations are much larger than unity. In our case, in steady state, the mode occupations $4\pi^2 n n_k$ range from about $10$ at $\kF$ to about $500$ for $k \rightarrow 0$.}\BibitemShut {Stop}%
\end{thebibliography}
\end{document}